\def\R{I \kern-.33em R}
\def\C{I \kern-.66em C}
\def\fs{\footnotesize}

\def\al{\alpha}
\def\bt{\beta}
\def\ga{\gamma}
\def\Ga{\Gamma}

\def\ka{\kappa}

\def\pt{\partial}
\def\na{\nabla}

\def\ka{K\"ahlerian\ }

\def\ba{\begin{eqnarray}}
\def\ea{\end{eqnarray}}

\documentstyle[11pt]{article}
\begin{document}

\title{ Anti -- \ka Manifolds }

\author{
Andrzej Borowiec\thanks{
E-mail: borow@ift.uni.wroc.pl}\\
{\fs Institute of Theoretical Physics}\\ {\fs University of Wroc{\l}aw}\\
{\fs pl. Maksa Borna 9, 50-204 WROC{\L}AW (POLAND)}\\ Mauro Francaviglia\\
{\fs Dipartimento di Matematica }\\ %
{\fs Universit\`a di Torino}\\ %
{\fs Via C. Alberto 10, 10123 TORINO (ITALY)}\\ Igor Volovich\thanks{
E-mail: volovich@mi.ras.ru}\\
{\fs Steklov Mathematical Institute}\\
{\fs Russian Academy of Sciences}\\
{\fs Gubkin St. 8, GSP--1, 117966 MOSCOW (RUSSIA)} }%
\date{\bf math-ph/9906012}
\maketitle

\begin{abstract}

An anti-\ka manifold is a complex manifold with an anti-Hermitian 
metric  and a parallel almost complex
structure. It is shown that a metric on such a manifold must be 
the  real part of a holomorphic
metric. It is proved that all odd Chern numbers of an
anti-\ka manifold vanish and that complex parallelisable manifolds 
(in particular the factor space $G/D$ of a complex Lie group $G$ 
over the discrete subgroup $D$ ) are anti-\ka manifolds. A method of
generating new solutions of Einstein equations by using the theory of
anti-\ka manifolds is presented. \end{abstract} 

\section{Introduction}

\ \ \
K\"ahlerian manifolds constitute a major class of Riemannian (complex)
manifolds and powerful methods of complex differential geometry have 
been developed to investigate their properties (see e.g. \cite{Bes}). 
In this paper we shall consider an apparently new class of 
pseudo-Riemannian manifolds
which will be called {\it anti-K\"ahlerian} manifolds and which are 
also deeply related with complex analysis.

Recall that a {\it \ka manifold}\ can be defined as a triple $(M, g, J)$
which consists of a smooth manifold $M$ endowed with an almost complex
structure $J$ and a Riemannian metric $g$ such that $\na J = 0$, where
$\na$ is the Levi-Civita connection of $g$ and the 
metric $g$ is assumed to
be Hermitian: $g(JX, JY)=g(X, Y)$ for all vectorfields $X$ and $Y$ on $M$.
By an {\it anti-\ka manifold} we mean instead a triple $(M, g, J)$ which
consists of a smooth manifold $M$, an almost complex structure $J$ and a
metric $g$ such that $\na J=0$, where $\na$ is the Levi-Civita connection
of $g$ and the metric $g$ is anti-Hermitian: $g(JX, JY)=-g(X, Y)$
for all vectorfields $X$ and $Y$ on $M$. An almost-complex anti-Hermitian
structure defines in particular an $O(m,\C)$-structure on $M$. Structures
of this kind have been also studied under the names: almost complex
structure with Norden (or $B$-) metric \cite{Nor,CHR,GI,BB,OP}.

In this paper anti-\ka manifolds will be investigated. In particular we 
will show that any such metric $g$ should be the real part of a 
certain holomorphic
metric on $M$. Holomorphic Riemannian metrics on "complex space-times" 
have been discussed in Pleba\'{n}ski \cite{Ple}, Penrose \cite{Pen}. 
Let us mention that the characterization of complex analytic Riemannian 
metrics in terms of complex connections has been considered in 
Ivanov \cite{Iva}.

We also consider the Chern numbers of anti-\ka manifolds and anti-\ka
Einstein manifolds. A family of examples of anti-\ka manifolds will be
given. It will be shown that all odd Chern numbers of anti-\ka manifolds 
vanish and that a compact \ka manifold cannot be anti-K\"ahlerian. It will 
also be shown that complex parallelisable manifolds (in particular the 
factor space $G/D$ of a complex Lie group $G$ over the discrete subgroup 
$D$ ) are anti-\ka manifolds.

Finally, we show that the complexification of a given Einstein metric 
leads to a method of generating new solutions of Einstein
equations from a given one.
In particular, it implies that the class of Einstein metrics with
neutral signature is the largest one in the following sense: any
Einstein metric, (of arbitrary signature) in dimension $m$, generates 
(at least locally) an infinite family of Einstein metrics with
neutral signatures in dimensions $2^k m$, $k=1,2,\ldots$ \ .
\newpage
Anti-\ka manifolds appeared naturally in a previous work of ours in
connection with a variational principle in nonlinear theories of gravity
\cite{BFFV2,BFFV3}.
\vskip1cm

\section{ Holomorphic metrics}

\ \ \
Let $(M,J)$ be a $2m$-dimensional real almost-complex manifold and let $g$ 
be an anti-Hermitian metric on $M$:
$$g(JX, JY)=-g(X, Y) 
$$
or equivalently:
$$g(JX, Y)=g(X, JY) 
$$
Then the metric $g$ has necessarily a neutral (Kleinian) signature $(m,m)$.
We extend $J,g$ and the Levi-Civita connection of $g$ in the well known 
way by $\C$-linearity to the complexification of the tangent bundle
$TM^{C}=TM\otimes \C$. Let us now fix a (real) basis
$\{X_1,...,X_m,JX_1,...,JX_m\} $ in each tangent space $T_xM$; then the set
$\{Z_a, Z_{\bar a}\}$, where $Z_a=X_a-iJX_a,\ Z_{\bar a}=X_a+iJX_a$, forms
a basis for each complexified tangent space $T_x M\otimes\C$. Unless
otherwise
stated, little Latin indices $a,b,c,...$ run from $1$ to $m$, while Latin
capitals $A,B,C,...$ run through $1,...,m,{\bar 1},...,{\bar m}$; for
notational convenience we shall also bar capital indices and we shall 
assume ${\bar{\bar A}}=A$. One has $JZ_a=iZ_a$ and $J Z_{\bar a}=
-i Z_{\bar a}$. We set $ g_{AB}=g( Z_A,Z_B)=g_{BA}$. Then the 
following holds:\smallskip\\

{\bf Proposition 2.1.}\ {\em Let $(M, J)$ be an almost-complex manifold and
$g$ be an anti-Hermitian metric on it. Then the complex extended metric $g$
(in the complex basis introduced above) satisfies the following conditions
\ba\label{b1}
g_{a\bar b}=g_{\bar ba}=0
\ea
\ba\label{b2}
g_{\bar A\bar B}={\bar g_{AB}}
\ea
Conversely, if the complex extended metric $g_{AB}$ satisfies (1-2) then
the initial metric must be anti-Hermitian.}

The proof of this result is straightforward.

\vskip1truecm

It will be customary, in this note, to write a metric satisfying (1-2) as
\ba\label{b3}
ds^2=g_{ab}dz^adz^b+g_{\bar a\bar b}dz^{\bar a}dz^{\bar b} \ea since in
adapted almost-complex coordinates (see below) \newline
$x^\mu=(x^a, y^a\equiv x^{m+a})$, $z^a=x^a+iy^a$ one has $$
g_{\mu\nu}dx^\mu dx^\nu = 2\ {\hbox{Re}}\ [g_{ab}dz^adz^b] $$ where
$\mu=1,\ldots , 2m$, $a=1,\ldots , m$ and $Re$ reads Real Part.

We define now the complex Christoffel symbols $\Ga^C_{AB}$ by \ba\label{b4}
\nabla_{Z_A}Z_B=\Ga^C_{AB}Z_C
\ea
It is known \cite{KN} that if $\nabla J=0$ then the torsion $T$ and the
Nijenhuis tensor $N$ satisfy the identity \ba\label{b5}
T(JX,JY)=\frac{1}{2}N(X,Y)
\ea
for all vectorfields $X$ and $Y$. Since the complex extended Levi-Civita
connection $\nabla$ has vanishing torsion,
the complex Christoffel symbols are symmetric, i.e.:
$\Ga^C_{AB}=\Ga^C_{BA}$.
In this case the complex structure $J$ is integrable, so that the real
manifold $M$ inherits the structure of
 a complex manifold. Let us now recall (see e.g. \cite{KN}) that there is a
one-to-one correspondence between complex manifolds and
real manifolds with an integrable complex structure. This means that there
exists an atlas of real, adapted (local) coordinates
$(x^1,...,x^m,y^1,...,y^m)$ such that $J(\partial/\partial
x^a)=\partial/\partial y^a,\ \ J(\partial/\partial
y^a)=-\partial/\partial x^a$. Setting then $z^a=x^a+iy^a$ and taking
$X_a=\partial/\partial x^a$ one gets \ba\label{b6}
Z_a=X_a-iJX_a=2\partial_a,\ \ Z_{\bar a}=X_a+iJX_a =2\partial_{\bar a} \ea
where $\partial_A=\partial/\partial z^A$ and $z^{\bar
a}={\bar z}^a$. It appears that $(z^a)$ form an atlas of complex (analytic)
coordinate charts on $M$. Now, by using Christoffel
formulae, one gets
$$\Ga^C_{AB}=\frac{1}{2}g^{CD}(Z_Ag_{BD}+Z_Bg_{DA}-Z_Dg_{AB}) $$
\ba\label{b7}=g^{CD}(\partial_Ag_{BD}+\partial_Bg_{DA}-\partial_Dg_{AB})
\ea Then the following holds (see also \cite{BFFV3,Iva}):
\vskip0.5truecm {\bf Theorem 2.2.}\ {\em Let $M$ be a $m$-dimensional
complex manifold, seen as a real $2m$-dimensional manifold
with a complex structure $J$. Let us further assume that $M$ is provided
with an anti-Hermitian metric $g$. We extend $J$, $g$ and
the Levi-Civita connection $\nabla$ by $\C$-linearity to the complexified
tangent bundle $TM^C$. Then the following conditions are
equivalent:

$(i)$
\ba\label{b8}
\nabla_{X}(JY)=J(\nabla_XY)
\ea
where $X$ and $Y$ are arbitrary real vectorfields;

$(ii)$
in all (local) complex coordinate systems $(z^1,...,z^m)$ on $M$ the
(complex) Christoffel symbols satisfy \ba\label{b9} \Ga ^C_{AB}=0\ \ \
\hbox{except} \ \hbox{for}\ \ \Ga_{ab}^c\ \hbox{ and}\ \ \Ga_{\bar
a\bar b}^{\bar c }={\bar \Ga_{ab}^c}
\ea

$(iii)$
in all (local) complex coordinate systems $(z^1,...,z^m)$ on $M$ the 
components of the complex extended metric $g_{ab}$ have the canonical 
form (3) and moreover they are
holomorphic functions, i.e. \ba\label{b10} 
\partial_{{\bar c}}\, g_{ab}=0
\ea}

{\it Proof.}\ From (\ref{b4}) we have
$${\bar \Ga}^C_{AB}=
\Ga^{\bar C}_{{\bar A}{\bar B}}
$$
The connection satisfies the conditions
$$\nabla_{Z_B}(JZ_c)=J\nabla_{Z_B}Z_c=i\nabla_{Z_B}(Z_c)$$
$$\nabla_{Z_B}(JZ_{\bar c})=
J\nabla_{Z_B}Z_{\bar c}= -i\nabla_{Z_B}(Z_{\bar c})$$ if and only if
\ba\label{b11}
\Ga^{ a}_{B{\bar c}}=\Ga^{\bar a}_{Bc}=0 \ea This proves the equivalence
between $(i)$ and $(ii)$. Then for the Christoffel symbols (\ref{b7}),
by taking (\ref{b1}) into account one gets \ba\label{b12}
\Ga^{ a}_{b{\bar c}}=g^{aD}(\partial_bg_{\bar cD}+\partial_{\bar c}g_{Db} 
- \partial_Dg_{b\bar c})=
g^{ad}\partial_{\bar c}g_{bd}
\ea
and from (\ref{b11}) it follows that
\ba\label{b13}
\partial_{\bar c}g_{bd}=0
\ea

Also the other relations (\ref{b11}) are reduced to (\ref{b13}) or its
complex conjugate.
Therefore the relation (\ref{b13}) is equivalent to (\ref{b10}). This
proves the equivalence
between
$(i)$ and $(iii)$. Our claim is thence proved. \hfill (Q.E.D.)\smallskip\\

\section{ Chern Classes of anti-\ka manifolds}

\ \ \
Here we consider some conditions on a manifold for the existence of an 
anti-\ka metric. Let $(M,J,g)$ be an anti-\ka manifold. Then $M$ is a 
complex manifold and
according to Theorem 2.2 there exists a holomorphic metric on $M$.
Therefore there is a complex
isomorphism between the complex tangent bundle $\tau$ and its dual
$\tau^*$. From the known
properties of the Chern classes $c_j(\tau^*)=(-1)^jc_j(\tau)$ one gets 
the following:

\vskip0.5truecm

{\bf Proposition 3.1.}\footnote{This proposition was suggested to us by
R.Narasimhan} \
{\em All odd Chern classes of an anti-\ka manifold $M$ vanish: $$
c_{2j+1}(M)=0 \ \ \ \ \ \ \forall j
$$}
\vskip0.5truecm
The following proposition shows that if a simply connected manifold is 
\ka then it cannot be anti-K\"ahlerian. \vskip0.5truecm

{\bf Proposition 3.2.} \
{\em If $M$ is a compact simply connected \ka manifold then it does not 
admit an anti-\ka metric.}

{\it Proof.} If $c_1(M)\neq 0$ then according to proposition 3.1 the
manifold $M$
cannot be anti-K\"ahlerian. Now if $c_1(M)=0$ and $M$ is a compact simply
connected \ka manifold then by a theorem due to Kobayashi \cite{Kob} we 
have $\Gamma(S^mTM)=\Gamma(S^mT^*M) =0$ for $m>0$ and therefore $M$ does 
not admit a holomorphic metric.
So it cannot be anti-K\"ahlerian.	\hfill (Q.E.D.)

\section{ Examples of anti-\ka manifolds}

\ \ \
Let us now consider the question:
{\it which manifolds may admit an anti-\ka structure?}

Let us first discuss compact manifolds. According to Proposition 3.1, in a
compact complex manifold $(M,J)$ which admits a holomorphic metric all odd
Chern numbers must vanish: $c_{2j+1} (M)=0$. In complex dimension 1 one
has $\chi (M)=c_1 (M)=0$, where $\chi (M)$ is the Euler cheracteristic of
$M$; therefore $M$ is a torus. On any torus one has a holomorphic metric
and the corresponding real anti-\ka metric will be Lorentzian.

In complex dimension 2, if $M$ is a regular (i.e., without holomorphic
1-forms) compact connected (complex) surface with vanishing first Chern
class, then it is a $K3$ surface; moreover it is known that any $K3$
surface is \ka \cite{Siu} and simply connected and therefore, in virtue of
Proposition 3.2, it does not admit a holomorphic metric. We have then to
consider irregular surfaces to find anti-\ka manifolds.

An interesting open question is whether the Hopf manifolds $S^{2p+1}\times
S^{2q+1}$ admit anti-\ka metrics.

Let us present now a large class of anti-\ka manifolds. A $m$-dimensional
complex manifold $M$ is called (complex) {\it parallelisable} if there exist
$m$ holomorphic vector fields {\bf $e_1,...,e_m$} which are everywhere
linearly independent in $M$. Every complex Lie group $G$ is parallelisable.
If $D$ is a discrete subgroup of $G$ then the complex manifold $G/D$ is also
parallelisable \cite{KN}. Conversely, Wang proved \cite{Wang,KN} that every
compact parallelisable manifold can be presented as the factor space $G/D$
of a complex Lie group $G$ over a discrete subgroup $D$. \vskip0.5truecm

{\bf Proposition 4.1.}\ {\em Every complex parallelisable manifold $M$ is an
anti-\ka manifold.}

{\it Proof}. Let us take a complex chart $(z^{\mu}), \ \ \mu =1,...,m$, on
$M$ and let
$(e_a^{\mu})$ be the components of the independent holomorphic vector fields
in this chart,$a=1,...,m$. Since the vectorfields are linearly independent
the inverse matrix $f_{\mu}^a$ defines $m$ holomorphic covectorfields
$e_a^{\mu}f_{\mu}^b =\delta^b_a$. Now let us set
$$g_{\mu\nu}=f_{\mu}^af_{\nu}^b\delta_{ab} $$ Then $g_{\mu\nu}$ is a
holomorphic metric on $M$. Therefore the manifold $M$ is
anti-K\"ahlerian.\hfill (Q.E.D.)

It would be interesting to know if every (compact) anti-\ka manifold is
complex parallelisable.

\vskip0.5truecm

{\it Remark.} It is known that between complex parallelisable manifolds
only the tori admit \ka metrics \cite{Wang}.

\section{ Anti K\"ahlerian Einstein manifolds} \ \ \
In this section we show that by taking the real part of a holomorphic
Einstein metric on a complex manifold of complex dimension $m$
one gets a real Einstein manifold of real dimension $2m$. Recall that a
metric $g$ is said to be an {\it Einstein metric} if $$
Ric(g)=\ga g	\eqno (14)
$$
where $\ga$ is a real constant and $Ric(g)$ denotes the Ricci tensor 
of the metric $g$, i.e.
$R_{\mu\nu}=R^\tau_{\mu\tau\nu}$, where $Riem(g)=R^\tau_{\mu\rho\nu}$ 
is the Riemann curvature tensor of the metric $g$.

The following theorem relating the real and complex Einstein equations 
holds true: \vskip0.5truecm

{\bf Theorem 5.1.}\ {\em Suppose that $(M,g,J)$ is an anti-K\"ahlerian
manifold, i.e. a complex manifold of complex dimension $m$ with a
holomorphic
metric $\hat g\equiv (g_{ab}(z)), a,b=1,...,m$ and a real metric $g\equiv
(g_{\mu\nu}(x))$, $\mu,\nu=1,...,2m$ defined by (3). Then the holomorphic
metric $g_{ab}(z)$ is Einstein if and only if the real metric
$g_{\mu\nu}(x)$ is a solution of the Einstein equations (14). In other
words we have: $Ric(g)=\ga g$ iff $Ric(\hat g)=\ga\hat g$, i.e. in 
components:
$$
R_{\mu\nu}(g)=\ga g_{\mu\nu}\ \ \ \hbox{iff}\ \ \ 
R_{ab}(\hat g)=\ga g_{ab}
$$
}

To prove Theorem 5.1 we need first the following lemma, whose proof is 
just a simple modification of the technical proof of Proposition 3.6 in 
\cite{KN} for the K\"ahlerian case.
\vskip0.5truecm

{\bf Lemma 5.2.}\ {\em The Riemann $Riem(g)$ and the Ricci $Ric(g)$ 
tensors of the (real) anti-Hermitian metric $g$ of an anti-K\"ahlerian 
manifold $(M,g,J)$ satisfy the conditions: 
$$
Riem(g)(X,Y)\circ J=J\circ Riem(g)(X, Y) \eqno(15a)$$ $$
Riem(g)(X, Y)= -\ Riem(g)(JX, JY) \eqno(15b) $$
$$
Riem(g)(JX, Y)= J\circ Riem(g)(X, Y) \eqno(15c)$$ $$
Ric(g)(JX,JY)=-\ Ric(g)(X,Y) \ \ \eqno (16) $$
for each $X,Y\in\chi(M)$.}

Notice that (15c) is a simple combination of (15a-b) and the first 
Bianchi identity (see also Lemma 1.1 in \cite{BB}). \vskip0.5truecm

We can now prove Theorem 5.1.
We shall not discuss here the Einstein equations for a generic metric 
of the form (3) but consider only the case when $g_{ab}$ is a 
holomorphic function.
From (10) and (15a-b) we get then for the Riemann tensor:
$$
R^D_{ABC}=0\ \ \ \hbox{except} \ \hbox{for}\ \ R^d_{abc}\ \hbox{ and}\ \
R^{\bar d}_{\bar a\bar b\bar c}={\bar R^d_{abc}} \eqno (17) $$
Moreover, the Christoffel symbols and Riemann tensor are given 
(in complex coordinates) by the classical formulae:
$$
\Ga_{ab}^c= \frac{1}{2} g^{cd}(\pt_a g_{bd}+\pt_b g_{ad} -\pt_d g_{ab})
\eqno (18a)$$ and
$$
R^a_{bcd} =\pt_c\Ga^a_{bd}-\pt_d\Ga^a_{bc} + \Ga^a_{ec}\Ga^e_{bd} -
\Ga^a_{ed}\Ga^e_{bc} \eqno (18b)
$$
(see also \cite{LeB} p.  174,
where complex-analytic self-dual Einstein metrics 
have been studied with some details).  Our aim is to establish a link
between the complex Ricci tensor $R_{ab}\equiv R^c_{acb}$ 
and the real one $R_{\mu\nu}\equiv R^\rho_{\mu\rho\nu}$. 

Define $\hat R(X, Y)V\equiv Riem(g)(V, X)Y$ then $Ric(g)(X, Y)=
\hbox{tr}\,\hat R(X, Y)$,
where $\hbox{tr}$ means the (real) trace of the $\R$-linear endomorphism
$\hat R(X,Y): T_xM\rightarrow T_xM$. Due to (15c) $J\circ\hat R(X,Y) =
\hat R(X,Y)\circ J$, i.e. $\hat R(X,Y)$ is $\C$-linear on $(T_xM, J)$.
It implies that  the trace of the endomorphism $\hat R(X,Y)$ 
(or its  $\C$-linear extension into $T^C_xM$) do satisfy
$$
\hat R(X, Y)^\mu_\mu \equiv \hat R(X, Y)^a_a + 
\hat R(X, Y)^{\bar a}_{\bar a} \equiv
2\ \hbox{Re}\ [\hat R(X, Y)^a_a] = 2\ \hat R(X, Y)^a_a
$$
Now, from (16) the complex Ricci tensor $R_{ab}$ is related with the real 
one $R_{\mu\nu}$ via Proposition 2.1 . In particular, analogously to 
(1) we have 
$$
R_{a\bar b}=0 
$$

The (complex) Einstein equations
$$
R_{AB}(g)=\ga g_{AB} 
$$
are thus equivalent to a pair of equations $$
R_{ab}(g_{cd})=\ga g_{ab} \eqno (19a)$$
$$
R_{\bar a\bar b}(g_{\bar c\bar d})=\ga g_{\bar a\bar b} \eqno (19b) $$

To get a real solution of Einstein equations (14) from (19) one uses 
then real coordinates $(x^{\mu}), \mu =1,...,2m$ on $M,$ i.e. 
$z^a=x^a+ix^{m+a}, a=1,...,m$ and writes the real 
Ricci tensor $R_{\mu\nu}$ in the form (3)
$$
R_{\mu\nu}dx^{\mu}dx^{\nu} =
R_{ab}dz^adz^b+R_{\bar a\bar b}dz^{\bar a}dz^{\bar b} 
$$
The result then follows.\hfill(Q.E.D.)\\
\vskip0.5truecm

{\bf Remark 5.3.}\ 
Recall from \cite{BFFV3} that beside the original metric $g$
one has to our disposal, on an anti-\ka manifold $(M,G,J)$, another 
real metric of neutral signature so called {\it twin metric}: 
$h(X,Y)\equiv g(JX,Y)$. One finds
$$
h_{\mu\nu}dx^\mu dx^\nu = -2\ {\hbox{Im}}\ [g_{ab}dz^adz^b] 
\ \ \ \hbox{and}\ \ \ h_{ab}= i g_{ab} \eqno (20)$$ 
Since $\na^gJ=0$ and $J^\mu_\nu = h^{\mu\al}g_{\al\nu}$ both metrics 
have the same (real and complex) Christoffel symbols: $\na^g=\na^h$,
thus the same (real and complex) Riemann and Ricci tensors. In the real
case only one of two twin metrics can be Einstenian. In complex 
coordinates $R_{ab}(g_{cd})=\ga g_{ab}$ implies
$R_{ab}(h_{cd})= -i \ga h_{ab}$ , i.e. both holomorphic metrics 
are Einstein metrics at the same time. One can say, in this way, that 
the metric $h$ is  an Einstein metric with an imaginary cosmological 
constant.\\\vskip0.5truecm

The last Theorem can be used to construct a "tower" of solutions of
Einstein equations.
Let be given a real analytic $n$-dimensional Einstein manifold $M^n$ with
the Einstein metric $g_{\al\bt}(x): R_{\al\bt}(g)=\gamma
g_{\al\bt}$. Let $\hat {M}^{2n}$ be a certain complex analytic 
extension (complexification) of the manifold $M^n$ ($x^\al\mapsto
z^a=x^\al+iy^\al;\ \al , a=1,\ldots ,n$) with the complex analytic 
metric $\hat {g}_{ab}(z)$ which is an analytic continuation of the 
original metric $g_{\al\bt}(x)$ (see \cite{Shu,Man,Fla}). 
The pair $(\hat {M}^{2n}, \hat g)$
is an anti-\ka manifold since $\hat g$ is automatically analytic.
Moreover, one has the (complex) Einstein equation $R_{ab}(\hat g)=\gamma
\hat {g}_{ab}$ where $R_{ab}(\hat g)$ is
obtained from $\hat {g}_{ab}(z)$ by using the standard formulae for the
Ricci tensor with partial derivatives with respect to
the complex coordinates $z^a$ (see also (18) and \cite{LeB} p.174). 
This is so because all steps in the algorithm for calculating the Ricci 
tensor (e.g., taking the inverse metric, computing partial derivatives,
multiplications, etc..) do commute with the operation of analytic 
continuation $g(x)\mapsto\hat g(z)$. Now, by taking the real part 
of $\hat {g}_{ab}dz^adz^b$ one gets a new analytic (real) metric of 
neutral signature on the $2n$-dimensional real manifold $\hat M^{2n}$. 
On virtue of Theorem 5.1 this new metric is again Einsteinian. We can
continue in this way and get a $4n$- dimensional real analytic Einstein
metric, then the $8n$-dimensional and so on.

To produce a whole family of examples one can take the complex analytic
continuations of all real analytic solutions of Einstein equations. We 
give here a simple concrete example. Take the standard Einstein metric on 
the $m$-dimensional sphere $S^m\subset\R^{m+1}$. After analytic 
continuation one gets:
$$
ds^2=dz^adz^a+\frac{(z^adz^a)^2}{1-z^az^a}+\ \hbox{ complex conj.}=
g_{\mu\nu}dx^{\mu}dx^{\nu} \eqno (21) $$
This metric $ g_{\mu\nu}$ lives on "the complex sphere" $S_C^m$
($w_1^2+...+w_{m+1}^2=1$), which can be interpreted as a quadric
$\zeta_1^2+...+\zeta^2_{m+1}-\zeta_{m+2}^2=0$ in $\C P^{m+1}$ if one 
takes $w_i=\zeta_i/\zeta_{m+2}$. It gives a solution
of the Einstein equations (14) and provides an example of an 
anti - Hermitian Einstein manifold $(M,g,J).$ As a real manifold the 
complex sphere $S_C^m$ is diffeomorphic to the tangent bundle $TS^m$. 
In particular for $m=2$ we get a \underline{real} solution of Einstein 
equations on the 4-dimensional, non-compact real manifold $TS^2$ 
with a metric of Kleinian signature $(++--)$.

Notice also that any Einstein metric
on a compact Riemannian manifold $M^n$ leads to an anti-K\"ahlerian
Einstein metric on another real manifold $\hat M^{2n}$. It follows from
known facts \cite{Bou} that any Einstein metric is analytic in a certain
atlas on $M^n$. Therefore there exists a complex analytic continuation of
the metric to a complex manifold of complex dimension $n$ which is a real
anti-K\"ahlerian manifold $\hat M^{2n}$.

\section*{Acknowledgments}
\ \ \
\indent We are grateful to G. Alekseev, M. O. Katanaev, R. Narasimhan and
Z. Olszak for useful discussion. One of us (A.B.) is supported by Polish
KBN and Mexican CONACyT (\#27670 E). I.V. is supported by INTAS grant
960698.

\end{document}